\documentclass[twoside]{article}

\usepackage{graphicx}

\usepackage{cmpj2e}

\usepackage{amssymb,amsbsy,amsmath,color}

\newcommand{\bra}[1]{\ensuremath{\langle #1|}}
\newcommand{\ket}[1]{\ensuremath{|#1\rangle}}
\newcommand{\braket}[2]{\ensuremath{\langle #1|#2\rangle}}

\newcommand{\op}[1]{%
    \fontdimen12\textfont3=2pt\fontdimen12\scriptfont3=1.4pt%
    \!\null\mathop{\vphantom{#1}\smash{#1}}\limits_{\sim}\null\!}
\newcommand{\vecop}[1]{%
    \fontdimen12\textfont3=2pt\fontdimen12\scriptfont3=1.4pt%
    \!\null\mathop{\textbf{\vphantom{#1}\smash{#1}}}\limits_{\sim}\null\!}

\newcommand{\fmref}[1]{(\protect\ref{#1})}

\newtheorem{k-rule}{k-rule}

\newcommand{\mofe}[1]{\{$\textrm{Mo}_{72}\textrm{Fe}_{30}$\}}
\newcommand{\mocr}[1]{\{$\textrm{Mo}_{72}\textrm{Cr}_{30}$\}}
\newcommand{\mov}[1]{\{$\textrm{Mo}_{72}\textrm{V}_{30}$\}}
\newcommand{\cula}[1]{\{$\textrm{Cu}_{12}\textrm{La}_{8}$\}}

\newcommand{\reduced}[3]{\ensuremath{\langle #1||#2 || #3 \rangle}}

\newcommand{\com}[2]{\ensuremath{\left[ #1,\,#2 \right]}}

\newcommand{\threej}[6]{\ensuremath{\begin{pmatrix} &#1& \quad &#2& \quad &#3& \\ &#4& \quad &#5& \quad &#6& \end{pmatrix}}}

\newcommand{\ito}[3]{\ensuremath{\op{#1}^{(#2)}_{#3}}}
\newcommand{\vecito}[2]{\ensuremath{\vecop{#1}^{(#2)}}}

\title[]
{Approximate eigenvalue determination of geometrically frustrated magnetic molecules%
\thanks{rschnall@uos.de}}
\author[]{Roman Schnalle\refaddr{label2}, Andreas
  M. L{\"a}uchli\refaddr{label3}, 
  J{\"u}rgen Schnack\refaddr{label1}}
\addresses{
\addr{label2} Universit{\"a}t Osnabr{\"u}ck, Fachbereich Physik, D-49069 Osnabr{\"u}ck, Germany
\addr{label3} Max Planck Institut f{\"u}r Physik komplexer Systeme, N{\"o}thnitzerstr. 38, D-01187 Dresden, Germany
\addr{label1} Universit{\"a}t Bielefeld, Fakult{\"a}t f{\"u}r Physik, Postfach 100131, D-33501 Bielefeld, Germany
}
\begin{document}

\maketitle

\begin{abstract}
Geometrically frustrated magnetic molecules have attracted a lot
of interest in the field of molecular magnetism as well as
frustrated Heisenberg antiferromagnets. In this article we
demonstrate how an approximate diagonalization scheme can be
used in order to obtain thermodynamic and spectroscopic
information about frustrated magnetic molecules. To this end we
theoretically investigate an antiferromagnetically coupled spin
system with cuboctahedral structure modeled by an isotropic
Heisenberg Hamiltonian.
\pacs 75.10.Jm,75.50.Xx,75.40.Mg,75.50.Ee
\end{abstract}

\section{Introduction}

The complete understanding of small magnetic systems such as
magnetic molecules is compulsorily connected to the knowledge of
their energy spectra. From the energy spectra all spectroscopic,
dynamic, and thermodynamic properties of the spin systems can be
obtained. Unfortunately, an exact calculation of the spectrum is
often restricted due to the huge dimension of the Hilbert space
even if one works within the most simple isotropic Heisenberg
model. The dimension grows for a system of $N$ spins with spin
quantum number $s$ exponentially and is $(2s+1)^N$.

In order to get insight into the properties of large magnetic
molecules one can access several numerical methods which have
been developed in the past. Of course, the ultimate method of
choice would be an exact numerical diagonalization yielding the
complete energy spectrum. In recent years there has been
enormous progress on extending the range of applicability of the
exact numerical diagonalization of the Heisenberg model. To this
end the use of spin-rotational symmetry
\cite{GaP:GCI93,BCC:IC99} in combination with point-group
symmetries \cite{Wal:PRB00,BOS:TMP06,SBO:JPA07,ScS:PRB09} can be
of great advantage with respect to a reduction of computational
requirements, i.e. a need of hardware resources and computation
time. Apart from the exact numerical diagonalization technique
the magnetism of magnetic molecules can be very well
investigated using complementary methods such as Density Matrix
Renormalization Group (DMRG)
\cite{Whi:PRB93,ExS:PRB03,Sch:RMP05}, Lanczos
\cite{Lan:JRNBS50}, or Quantum Monte Carlo (QMC)
\cite{SaK:PRB91,San:PRB99,EnL:PRB06} techniques. Nevertheless,
also these methods suffer from theoretical limitations, QMC for
instance in systems with geometric frustration.

Currently magnetic molecules which exhibit geometric frustration
are of special interest due to the richness of physical
phenomena like plateaus and jumps of the magnetization for
varying field as well as special features of their spectra such
as low-lying singlets
\cite{SHS:PRL02,SSS:PRL05,SSR:JMMM05,SSR:PRB07}. In this respect
a lot of insight has been obtained by investigating molecular
representations of archimedean-type spin systems
\cite{GNZ:PRB06,RLM:PRB08}, i.e. systems in which participating
spins occupy the vertices of an archimedean solid. Such
representations have already been synthesized several years ago
and exist for example as \cula{} \cite{BGG:JCSDT97}
(cuboctahedron, $s=1/2$) and \mofe{} \cite{MSS:ACIE99,MLS:CPC01}
(icosidodecahedron, $s=5/2$). In this context the molecular
compound \mofe{} is probably one of the most investigated
magnetic molecules. However, a theoretical explanation of its
special physical properties was so far given mostly by
considering purely classical models \cite{AxL:PRB01,SPK:PRB08}
or directly related quantum mechanical counterparts like the
rotational-band model \cite{ScL:PRB00,SLM:EPL01}.

In this paper we want to show how the approximate
diagonalization technique which has been developed and applied
to unfrustrated, i.e. bipartite, magnetic molecules in
Ref. \cite{ScS:PRB09} can be used in order to determine the
energy spectrum of geometrically frustrated magnetic
molecules. The idea of this technique is to diagonalize the full
Hamiltonian in a reduced basis set. The basis set itself is an
eigenbasis of the rotational-band Hamiltonian. Such an ansatz
was already used by Oliver Waldmann \cite{WAL:PRB07} in order to
interpret inelastic neutron spectra of \mofe{}
\cite{GNZ:PRB06}. We will demonstrate that in contrast to
bipartite systems for a frustrated spin systems the rotational-band
states of all non-trivially different sublattice colorings
have to be taken into account in order to achieve a reliable
convergence of energy levels. Throughout this paper we use a
spin system with cuboctahedral structure and spin quantum
numbers $s=1$ and $s=3/2$ as well as an antiferromagnetic
coupling as an archetypical example of a frustrated magnetic
spin system.

This paper is organized as follows. In Sec. \ref{sec-2} the
general theoretical description of the system within the
Heisenberg model as well as general remarks on the use of
irreducible tensor operators and point-group symmetries are
given. In Sec. \ref{sec-3} the theoretical basics of the
approximate diagonalization within the isotropic Heisenberg
model are briefly reviewed and specified for the cuboctahedral
spin system. The convergence behaviour of the approximate
diagonalization is displayed and deeply discussed as well as the
specific heat and zero-field magnetic susceptibility for a
cuboctahedron with $s=3/2$. Furthermore, a numerically based
finding of an approximate selection rule is reported. This paper
closes with a Summary in Sec. \ref{sec-4}.

\section{Theoretical method} \label{sec-2}

In order to model the physics of antiferromagnetic molecules it
has been shown that an isotropic Heisenberg Hamiltonian with an
additional Zeeman-term and an antiferromagnetic nearest-neighbor
coupling provides the dominant terms. Such a Hamiltonian looks
like
\begin{equation} \label{eq:Hamiltonoperator}
   \op{H}= - \sum_{<i,j>} J_{ij} \vecop{s}(i) \cdot \vecop{s}(j)
   + g \mu_B \vecop{S} \cdot \vec{B}
\ .
\end{equation}
The indices of the sum are running over all pairs $<i,j>$ of
interacting spins $i$ and $j$. The first part consisting of the
sum over single spin operators $\vecop{s}(i)$ at sites $i$
interacting with the coupling strength $J_{ij}<0$ refers to the
Heisenberg exchange whereas the second part -- the Zeeman-term
-- couples the total spin $\vecop{S}$ to an external magnetic
field $\vec{B}$.

Taking without loss of generality $J_{ij}=J$ for interacting
spins the Heisenberg part assumes the following form
\begin{equation} \label{eq:Heisenberg_WW}
   \op{H}_\text{Heisenberg} = -J \sum_{<i,j>} \vecop{s}(i) \cdot \vecop{s}(j),
\end{equation}
where each coupling is counted only once. Since due to $SU(2)$
symmetry the commutation relations
$\com{\op{H}_\text{Heisenberg}}{\vecop{S}} = 0$ hold, a common
eigenbasis $\{ \ket{\nu} \}$ of $\op{H}_\text{Heisenberg}$,
$\vecop{S}^2$ and $\op{S}_z$ can be found and the effect of an
external magnetic field $\vec{B}=B \cdot \vec{e}_z$ can be
included later, i.e.
\begin{equation}
   E_\nu(B)=E_\nu + g \mu_B B M_\nu
\ .
\end{equation}
Here $E_\nu$ denotes the energy eigenvalues, $\ket{\nu}$ the
eigenstates, and $M_\nu$ denotes the corresponding magnetic
quantum number.

For the matrix representation of the Heisenberg Hamiltonian
\fmref{eq:Heisenberg_WW} the irreducible tensor operator
technique is used \cite{GaP:GCI93,BCC:IC99,ScS:PRB09}. The
Heisenberg Hamiltonian is expressed in terms of irreducible
tensor operators and its matrix elements are evaluated using the
Wigner-Eckart-theorem
\begin{eqnarray} \label{eq:wigner-eckart}
  && \bra{\alpha \, S \, M} \ito{T}{k}{q} \ket{\alpha' \, S' \, M'} = \nonumber \\
  && (-1)^{S-M} \reduced{\alpha \, S}{\vecito{T}{k}}{\alpha' \,
  S'} \threej{S}{k}{S'}{-M}{q}{M'}
\ .
\end{eqnarray}
Equation \fmref{eq:wigner-eckart} states that a matrix element
of the $q$-th component of an irreducible tensor operator
$\vecito{T}{k}$ of rank $k$ is given by the reduced matrix
element $\reduced{\alpha \, S}{\vecito{T}{k}}{\alpha' \, S'}$
and a factor containing a Wigner-3J symbol
\cite{VMK:quantum_theory}. The basis in which the Hamilton
matrix is set up is of the form $\ket{\alpha \, S \,
M}$. $\alpha$ refers to a set of intermediate quantum numbers
given by addition rules when coupling single spins
$\vecop{s}(i)$ to a total spin $\vecop{S}$ with spin quantum
number $S$. The underlying spin-coupling scheme directly
influences the form of the irreducible tensor operator
$\vecito{T}{k}$ and further on the successive calculation process
of the reduced matrix elements in Eq. \fmref{eq:wigner-eckart}.

By using irreducible tensor operators and the
Wigner-Eckart-theorem it is possible to drastically reduce the
dimensionality of the problem, i.e. of the Hamilton matrices
which have to be diagonalized numerically. The total Hilbert
space $\mathcal{H}$ can be decomposed into subspaces
$\mathcal{H}(S, M=S)$. Such a decomposition results in block
factorizing the Hamilton matrix where each block can be labelled
by the total spin quantum number $S$.

Additionally point-group symmetries lead to a further reduction
of the matrices. Apart from that, states are labelled by the
irreducible representation they are belonging to. This is very
helpful in understanding the physics of the system. The
incorporation of these symmetries results in symmetrized basis
states which are constructed by the projection operator
\cite{Tin:Group_theory}
\begin{equation} \label{eq:Projektionsoperator}
   \mathcal{P}^{(n)}\ket{\alpha \, S \, M} =
\left( \frac{l_n}{h} \sum_{R} \left(\chi^{(n)}(R)\right)^\ast
\, \op{G}(R) \right) \ket{\alpha \, S \, M}
\ .
\end{equation}
Here $l_n$ denotes the dimension of the $n$-th irreducible
representation of the point-group $\mathcal{G}$ which is of
order $h$. $\op{G}(R)$ refers to the symmetry operations of
$\mathcal{G}$ and $\chi^{(n)}(R)$ denotes its character with
respect to $n$. The effect of the symmetry operation $\op{G}(R)$
on basis states of the form $\ket{\alpha \, S \, M}$ is
discussed in Refs. \cite{Wal:PRB00,SBO:JPA07,ScS:PRB09}.

\section{Approximate diagonalization for frustrated systems: the
  cuboctahedron} \label{sec-3}

\begin{figure}[ht!]
\centering
\includegraphics*[clip,width=40mm]{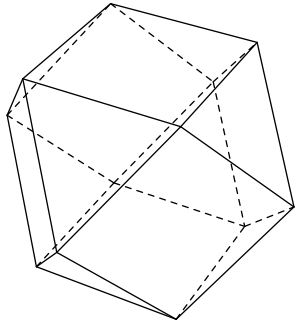}
\caption{Geometrical structure of the cuboctahedron \cite{mathworld}.}
\label{F-1}
\end{figure}

In this section we follow a method of calculating approximate
eigenstates and eigenvalues of the Heisenberg Hamiltonian in
Eq. \fmref{eq:Heisenberg_WW} and determine approximately the
energy spectrum of a spin system with cuboctahedral
structure. In this system $12$ spins of spin quantum number $s$
occupy the vertices of a cuboctahedron interacting along its
edges. The geometrical structure of the cuboctahedron is shown
in Fig. \ref{F-1}. A detailed description of the approximation
scheme is given in Ref. \cite{ScS:PRB09} where it was
successfully applied to determine approximate energy spectra and
thermodynamic properties of spin rings, i.e. bipartite
systems. This approximation rests on the idea to diagonalize the
full Hamiltonian in a reduced basis set. The basis set itself is
an eigenbasis of the rotational-band Hamiltonian which from the
point of view of perturbation theory can be understood as an
approximation to the full Hamiltonian. For bipartite systems
this (zeroth order) approximation is already very good
\cite{ScL:PRB00}. 

The Heisenberg Hamiltonian can be decomposed into two parts like
\begin{equation} \label{eq:approxDiag_gen}
   \op{H} = \op{H}_{\text{RB}} + \op{H}'
\ ,
\end{equation}
where $\op{H}_{\text{RB}}$ is the rotational-band Hamiltonian
\cite{ScL:PRB00,SLM:EPL01,Wal:PRB01} and $\op{H}'$ is an
operator containing the remaining terms. The rotational-band Hamiltonian
which is an effective quantum mechanical Hamiltonian based on
classical assumptions looks like
\begin{equation} \label{eq:H_rb_allgemein}
   \op{H}_\text{RB} = -\frac{D J}{2N} \left[\vecop{S}^2 - \sum_
   {n=1}^{N_s} \vecop{S}^2_n \right]
\ . 
\end{equation}
Here $N$ denotes the number of spins within the system and $N_s$
the number of sublattices the classical ground state of the
system is composed of. The prefactor $-DJ/(2N)$ can be seen as
the effective coupling strength between the total spin
$\vecop{S}$ and the sublattice spins $\vecop{S}_n$ which arise
from coupling all single spins $\vecop{s}(i)$ belonging to the
$n$-th sublattice. The value of $D$ directly depends on the
system. It is chosen such that the energy of the ferromagnetic
state of the system is matched; for the cuboctahedron it is
$D=6$.

The full Heisenberg Hamiltonian is now diagonalized within a
reduced set $\{\ket{\phi_i}\}$, $i=1,\dots,n_\text{red}$, of
basis states of $\op{H}_\text{RB}$ yielding approximate
eigenstates and eigenvalues of $\op{H}$. The set of approximate
basis states is energetically ordered.

\begin{figure}[ht!] 
   \centering
   \includegraphics*[width=70mm]{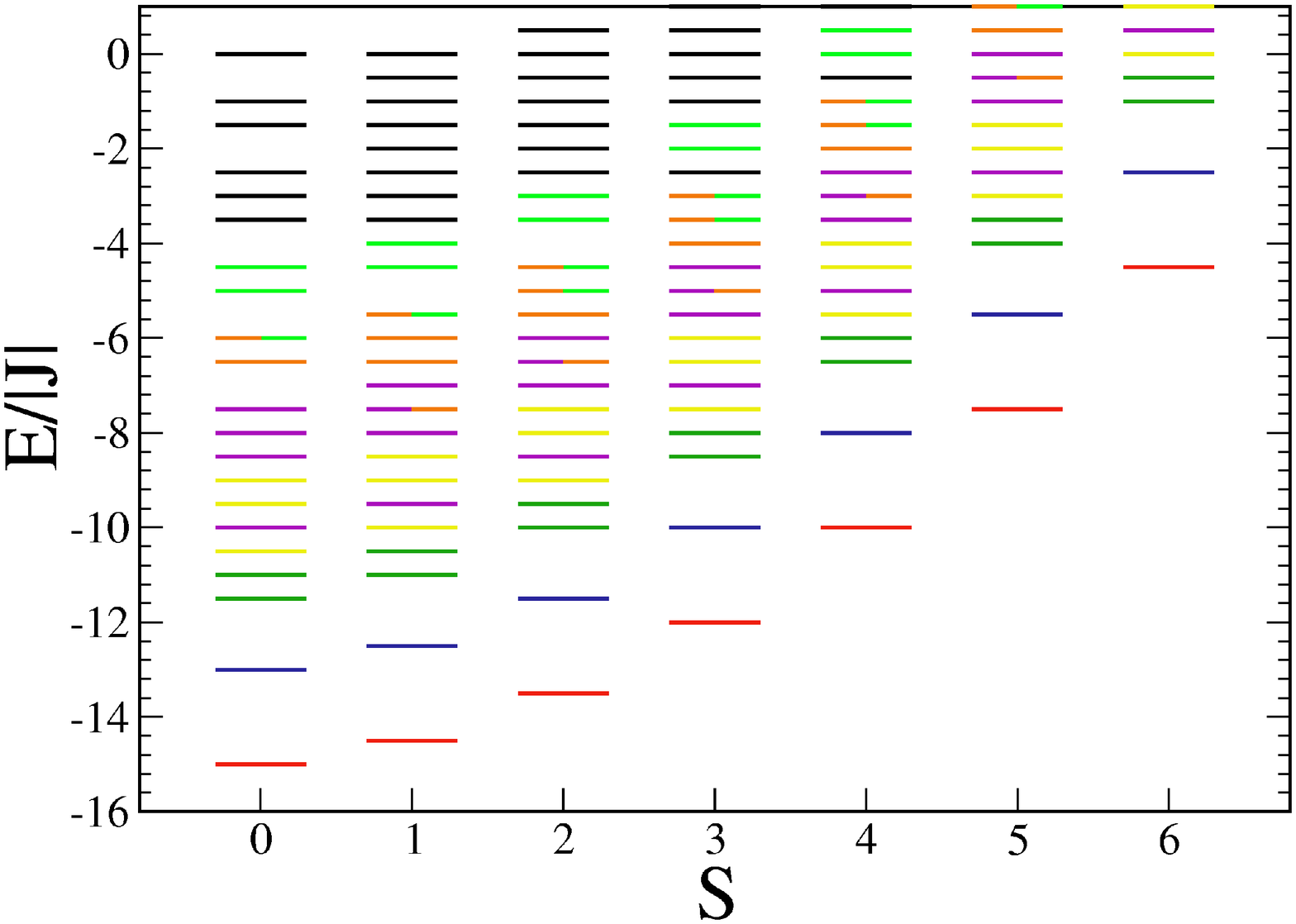}\hfill
   \includegraphics*[width=70mm]{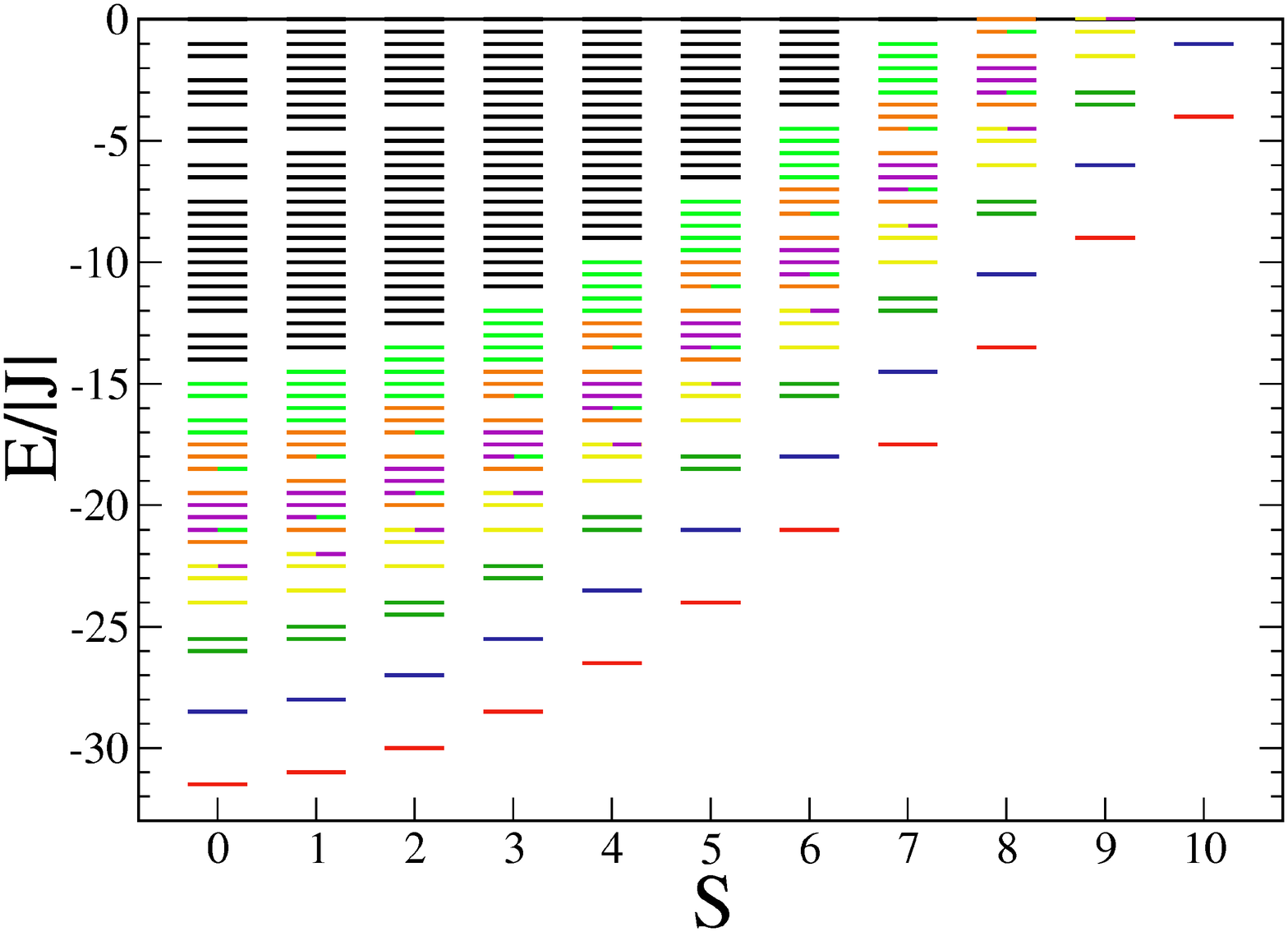}
   \caption{Part of the energy spectra of the rotational-band
   Hamiltonian for a cuboctahedron with twelve spins $s=1$
   (left) and $s=3/2$ (right). Seven super-bands are colored.} 
   \label{F-2}
\end{figure}

Before discussing the results of our approximate diagonalization
we would like to characterize the eigenbasis of
$\op{H}_\text{RB}$. Figure \ref{F-2} shows the low-lying part of
the energy spectra of $\op{H}_\text{RB}$ for a cuboctahedron
with $s=1$ and $s=3/2$ given by the rotational-band model. They
exclusively consist of parabolas -- so-called
rotational bands. The eigenvalues $E_\text{RB}(S_1,S_2,S_3,S)$
of $\op{H}_\text{RB}$ depend only on the spin quantum numbers of
the sublattice spins $S_n$, with $n=1, 2, 3$, and of the total
spin $S$. Corresponding eigenstates are trivial and analytically
given as $\ket{\alpha \, S_1 \, S_2 \, S_3 \, S \, M}$. The
additional quantum number $\alpha$ refers to a set of
intermediate spin quantum numbers which appears when coupling
single spins $\vecop{s}_i$ of the same sublattice to the
corresponding sublattice spins $\vecop{S}_i$ and further on
coupling the sublattice spins to the total spin
$\vecop{S}$. With regard to the set of intermediate spin quantum
numbers $\alpha$ the number of all possible ways of constructing
states characterized by fixed values of the sublattice and total
spin quantum numbers determine the degeneracy of energy levels
in the spectrum of the rotational-band Hamiltonian.

In the case of $s=1$ the lowest band in Fig. \ref{F-2} is given
by states $\ket{\alpha \, S_1 \, S_2 \, S_3 \, S \, M}$ with
sublattice spin quantum numbers $S_1=S_2=S_3=S_\text{max}=4
\cdot 1=4$ while the second band is given by a deviation of one
sublattice spin of 1, i.e. $\ket{\alpha \, (S_\text{max}-1) \,
S_\text{max} \, S_\text{max} \, S \, M}$ and permutations
thereof. The other bands can then be constructed by introducing
additional deviations of the sublattice spin quantum numbers.
The energy spectrum with $s=3/2$ shown in
Fig. \ref{F-2} can be constructed accordingly.

Additionally several energy levels in Fig. \ref{F-2} are
colored. This coloring refers to so-called super-bands. A
super-band consists of those rotational bands for which the sum
of sublattice spin quantum numbers is the same. In contrast to
bipartite systems (see Ref. \cite{ScS:PRB09}) the spectrum of
the rotational-band Hamiltonian for the cuboctahedron is much
denser at low energies and only the first three super-bands are
well separated from the others.

The classical ground state of the system plays the key role
within the approximate diagonalization. The better the quantum
mechanical system can be approximated by a classical picture of
the ground state the more effectively the approximate
diagonalization works, i.e. the faster the approximate energy
eigenvalues converge towards the exact values. From a purely
classical point of view the ground state of a cuboctahedron
exhibits a three-sublattice structure and is infinitely
degenerate since there coexist coplanar and non-coplanar vector
orientations \cite{ScL:JPA03}. Here it should be emphasized that
within the approximate diagonalization, i.e. within a quantum
mechanical treatment, coplanarity of the classical ground state
does not play a role, but the coloring of the classical ground
state does, since there is a direct impact on the set of
intermediate spin quantum numbers $\alpha$ in the
rotational-band states $\ket{\alpha \, S_1 \, S_2 \, S_3 \, S \,
M}$ which are taken into account as basis states in the
approximate diagonalization.

\begin{figure}[ht!] 
   \centering
   \includegraphics*[width=90mm]{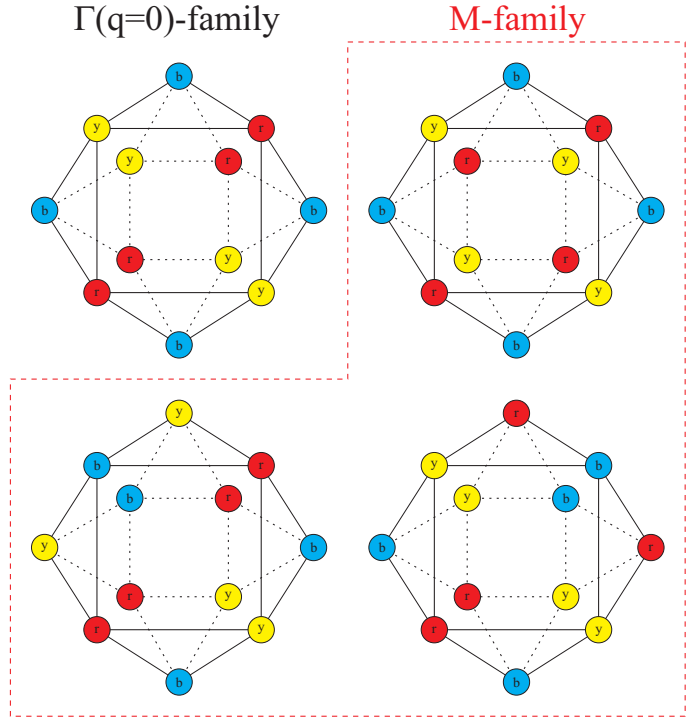}
   \caption{Families of the classical ground state of the
   cuboctahedron. Coloring of the $\Gamma(q=0)$-family (upper
   left). Coloring of the three equivalent states of the
   $M$-family (upper right and bottom). The drawing is a
	    schematic planar projection. The labels
   correspond to y: yellow, r: red, and b: blue.} 
   \label{F-3}
\end{figure}

The classical ground state of the cuboctahedron exhibits 24
colorings of the spins which can be -- by group theoretical
considerations -- decomposed into two families that are
invariant under operation of the full point-group symmetry $O_h$
\cite{RLM:PRB08}. Following Ref.~\cite{RLM:PRB08} these families
will be denoted as $\Gamma(q=0)$-family and $M$-family. It has
also been shown that those irreducible representations of $O_h$
which form the $\Gamma(q=0)$- and $M$-families are found in the
low-lying part of the spectrum of the quantum cuboctahedron with
half-integer spins $s=1/2,3/2,5/2$~\cite{RLM:PRB08}.
Figure \ref{F-3} shows the
colorings of the different classical ground state families of
the cuboctahedron. In a classical picture each color refers to a
sublattice with all spins pointing into the same direction. The
angle between classical spins belonging to different sublattices
is $120^\circ$.

In order to calculate the approximate spectrum of the system one
is now left with the construction of basis states of the form
$\ket{\alpha \, S_1 \, S_2 \, S_3 \, S \, M}$,
i.e. quasi-classical states. Therefore spins belonging to the
same sublattice have to be coupled to yield the total sublattice
spins $\vecop{S}_1$, $\vecop{S}_2$ and $\vecop{S}_3$. Afterwards
these total sublattice spins are coupled to the total spin
$\vecop{S}$. The underlying coupling scheme is given by a
classical ground state, i.e. a coloring from Fig. \ref{F-3} and
incorporated in the quantum number $\alpha$. In the following
the resulting basis states will be labelled with respect to the
classical reference state. To this end one has to distinguish
between basis states of the form $\ket{\gamma \, S_1 \, S_2 \,
S_3 \, S \, M}_\Gamma$, $\ket{\mu_1 \, S_1 \, S_2 \, S_3 \, S \,
M}_{M_1}$, $\ket{\mu_2 \, S_1 \, S_2 \, S_3 \, S \, M}_{M_2}$
and $\ket{\mu_3 \, S_1 \, S_2 \, S_3 \, S \, M}_{M_3}$. The
notation of the set of intermediate quantum numbers and the
subscript of the states now directly point to the classical
ground state colorings, i.e. the underlying coupling scheme. It
is important to note, that each of these four basis sets spans
the same Hilbert space $\mathcal{H}(S,M)$.

In the case of the $M$-family three different colorings
exist. When diagonalizing the Hamiltonian approximately while
additionally using point-group symmetries one has to restrict to
those symmetry groups where the symmetry operations do not alter
the sublattice structure. A symmetry operation has no impact on
the sublattice structure if the corresponding spin permutation
results in recoloring of spins where all spins of a given
sublattice maintain the same color, i.e. subscript (r: red, y:
yellow, b: blue). For example an operation which leads to a
cyclic permutation of colorings like
\begin{equation}
   r \rightarrow y \rightarrow b \rightarrow r
\end{equation}
has no impact on the sublattice structure.

While the sublattice structure of the $\Gamma(q=0)$-family is
left invariant under all symmetry operations of $O_h$, the
sublattice structure (i.e. coloring) of the $M$-family is
not. In order to restore the sublattice invariance of the
classical ground state belonging to the $M$-family the basis
states for the approximate diagonalization can be constructed
from a usually overcomplete set $M$ of basis states by an
orthonormalization procedure. This set consists of
rotational-band eigenstates from each coloring of the
$M$-family, i.e. is given by
\begin{equation} \label{eq:complete_set_M}
   M = \left\{ \ket{\mu_1 \, S_1 \, S_2 \, S_3 \, S \,
   M}^i_{M_1} \right\} + \left\{\ket{\mu_2 \, S_1 \, S_2 \, S_3
   \, S \, M}^i_{M_2} \right\} + \left\{\ket{\mu_3 \, S_1 \, S_2
   \, S_3 \, S \, M}^i_{M_3} \right\} 
   \ ,
\end{equation}
while the index $i$ is taking the values
$1,\dots,n_\text{red}$. Here $n_\text{red}$ reflects the overall
number of states contained in the incorporated rotational bands
which is independent of the choice of the coupling scheme.

Since the underlying coupling scheme is different regarding the
chosen coloring these states have to be converted into an -- in
general -- arbitrary reference scheme $A$ before calculations
can be performed. A transition between states belonging to
different coupling schemes can be calculated using general
recoupling coefficients \cite{FPV:CPC95,FPV:CPC97}. For the
transition of a basis state $\ket{\mu_1 \, S \, M}_{M_1}$
referring to a coloring $M_1$ of the $M$-family into the scheme
$A$ one yields
\begin{equation} \label{eq:basis_M}
   \ket{\alpha \, S \, M}_A = \sum_{\alpha,S_1,S_2,S_3} {_A}\braket{\alpha \, S \, M }{\mu_1 \, S \, M}_{M_1} \, \ket{\alpha \, S \, M}_{A}
   \ ,
\end{equation}
where the summation indices indicate that the summation is
running over all valid combinations of values for the
intermediate spin quantum numbers $\alpha$ as well as for the
sublattice spin quantum numbers. The transitions between states
of colorings $M_2$ and $M_3$ into the scheme $A$ can be obtained
analogously.

\begin{figure}[ht!] 
   \centering
   \includegraphics*[width=70mm]{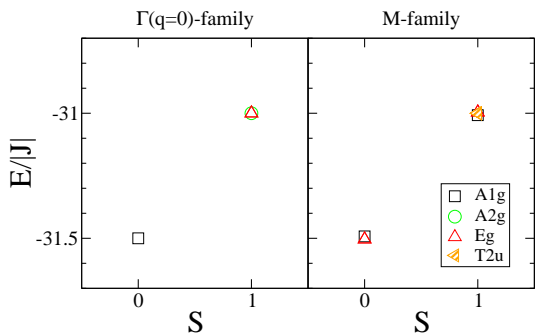}
   \caption{Classification by irreducible representations of
   $O_h$ of the quasi-classical states with lowest energies of
   the $\Gamma(q=0)$-family and $M$-family within $S=0$ and
   $S=1$ subspaces. The given energies refer to a
   diagonalization of the Heisenberg Hamiltonian exclusively
   within these quasi-classical states in the system with
   $s=3/2$.} 
   \label{F-4}
\end{figure}

Figure \ref{F-4} shows the classification of the quasi-classical
states with lowest energies within $S=0$ and $S=1$ subspaces
according to irreducible representations of $O_h$. It
corresponds to an approximate diagonalization of
$\op{H}_\text{Heisenberg}$ using only a reduced basis set of
states from the lowest rotational band in the system with
$s=3/2$. Obviously the classification of the quasi-classical
states is independent of the single spin quantum number
$s$. Solely, the energies are changed when approximately
diagonalizing the Hamiltonian of a system with different single
spin quantum numbers. Based on a classical ground state
belonging to the $\Gamma(q=0)$-family the states belong to the
irreducible representations $A_{1g}$ ($S=0$) and $A_{2g}$, $E_g$
(S=1). Taking the $M$-family as a starting point for the
construction of basis states the states belong to $A_{1g}$,
$E_g$ (S=0) and $A_{1g}$, $E_g$, $T_{2u}$ (S=1) where the
$T_{2u}$-state has apart from its intrinsic degeneration,
i.e. the dimension of the irreducible representation $T_{2u}$, a
twofold multiplicity.

The difference in the geometric properties of the lowest states
of the chosen ground-state family, expressed by their
decomposition into different irreducible representations, should
directly lead to a convergence behaviour which depends on the
choice of the underlying coloring. Since the approximate basis
set contains energetically ordered states (from lowest to
higher), low-lying states are expected to converge faster with
growing number of basis states used for the approximate
diagonalization \cite{ScS:PRB09}.

\subsection{Convergence of individual colorings} \label{sec-convergence}

\begin{figure}[ht!]
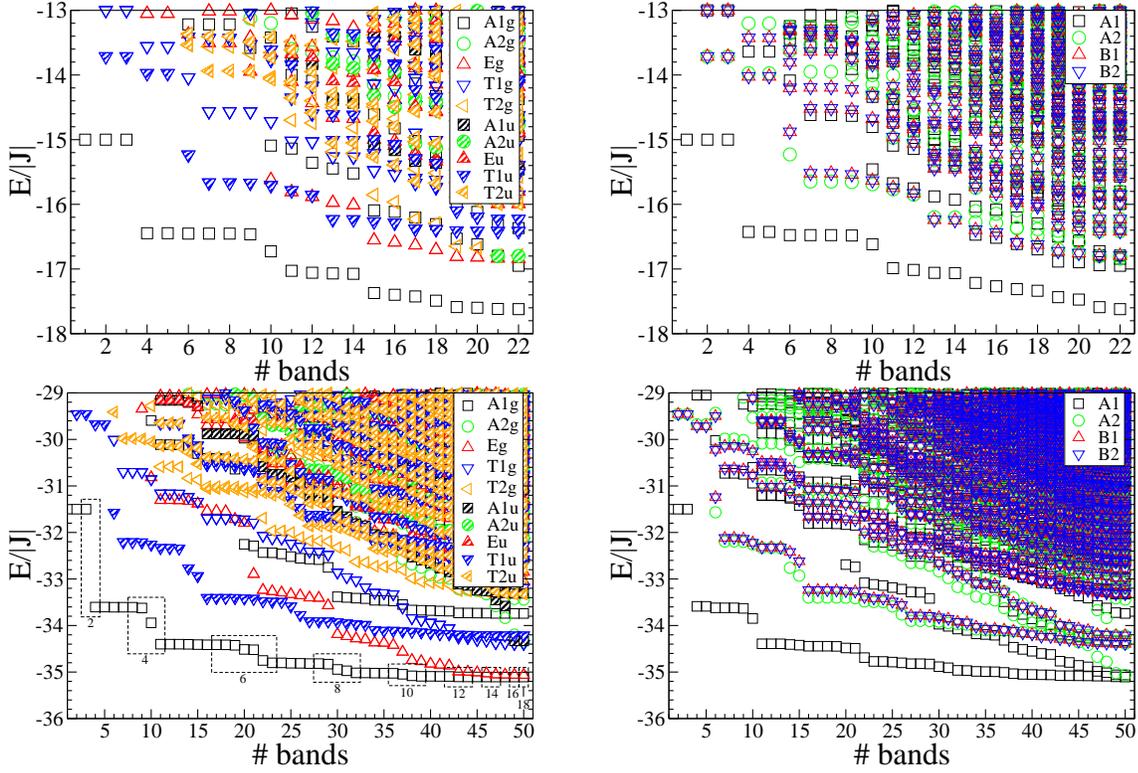
 
   \centering
   \includegraphics*[width=70mm]{ito-frus-fig-5a.eps} \hfill
   \includegraphics*[width=70mm]{ito-frus-fig-5b.eps} \\
   \includegraphics*[width=70mm]{ito-frus-fig-5c.eps} \hfill
   \includegraphics*[width=70mm]{ito-frus-fig-5d.eps}
   \caption{Low-lying energy spectra of a cuboctahedron with
   $s=1$ (upper row) and $s=3/2$ (lower row) within $S=0$
   subspaces using an increasing number of bands (states). The
   underlying classical ground states belong to the
   $\Gamma(q=0)$-family (left) and to one coloring of the
   $M$-family (right) while $O_h$ (left) and $D_2$ (right)
   point-group symmetries were used in order to classify the
   states. The dashed boxes in the lower left figure indicate
   regions, i.e super-bands, in which the energy of the ground
   state is considerably lowered. The numbers refer to the
   number of magnons of the super-bands.} 
   \label{F-5}
\end{figure}

In Fig. \ref{F-5} the approximate eigenvalues of a cuboctahedron
with $s=1$ and $s=3/2$ in dependence on the number of
incorporated bands within the $S=0$ subspaces are shown. As
underlying classical ground states for each system the
$\Gamma(q=0)$-family as well as one member of the $M$-family
have been chosen. In every subfigure the last column refers to
results from a complete diagonalization within
$\mathcal{H}(S=0)$. Since the $\Gamma(q=0)$-family is invariant
under $O_h$, the full point-group symmetry of a cuboctahedron
($O_h$) was used in order to classify the states. As mentioned
before the individual members of the $M$-family are not
invariant under $O_h$ but under $D_2$. Thus, in this case the
$D_2$ point-group symmetry was applied.

Looking at Fig.~\ref{F-5} it becomes apparent that the
convergence is directly dependent on the choice of the
underlying classical ground states. In agreement with
theoretical expectations states which correspond to low-lying
eigenstates of the rotational-band model converge faster than
higher-lying states. Additionally, by comparing the spectra of
$s=1$ and $s=3/2$ it can be seen that the ground state energy
converges more rapidly with increasing single spin quantum
number $s$. Nevertheless, apart from some states which exhibit a
quite regular and fast convergence behaviour the convergence is
-- overall -- rather poor. Especially the so-called low-lying
singlet\footnote{Here: the lowest $E_g$-state in the system with
$s=3/2$ based on a classical ground state coloring of the
$\Gamma(q=0)$-family.}, which appears in the case of
half-integer spins $s$ below the first triplet contributions,
converges very slowly. This means that it is poorly approximated
by one or a few eigenstates of the rotational-band Hamiltonian,
but instead is a superposition of very many basis states.

One significant property which becomes evident when tracing
single energy levels is that the convergence is stepwise -- at
least in the low-lying part of the spectra. Regions in which the
ground state is considerably lowered are exemplarily marked
with dashed boxes in the case of a cuboctahedron $s=3/2$
assuming a classical reference state of the
$\Gamma(q=0)$-family. The numbers below the boxes refer to the
number of magnons existing in the states of those
rotational bands which are incorporated within the marked
regions. Obviously the ground state energy is lowered whenever
super-bands with an even number of magnons are incorporated in
the approximate diagonalization. The observation of a stepwise
convergence leads to a helpful additional approximation given by
an approximate selection rule which is discussed below (see
Sec. \ref{sec-selection}). It should be mentioned here that
qualitatively the same convergence behaviour can be observed
within subspaces of $S > 0$. However, additional information
cannot be extracted from graphical visualizations of the
convergence in these subspaces, thus they will not be presented
here.

\subsection{Convergence of combined colorings}

In order to improve the convergence of the energy levels in
comparison to the behaviour shown in Fig.~\ref{F-5} the
decomposition of the quasi-classical states in Fig. \ref{F-4}
directly leads to the starting point. Since the quasi-classical
states, that result from different colorings, can be classified
according to irreducible representations of $2 A_{1g}$ and $E_g$,
it is a straightforward task to use linear combinations of
rotational-band eigenstates of the $\Gamma(q=0)$-family as well
as of the $M$-family as basis states for the approximate
diagonalization. The set of basis states results from an
extension of $M$ in Eq. \fmref{eq:complete_set_M} to
\begin{equation} \begin{split} \label{eq:complete_set_all}
   G &= \left\{ \ket{\mu_1 \, S_1 \, S_2 \, S_3 \, S \,
   M}^i_{M_1} \right\} + \left\{\ket{\mu_2 \, S_1 \, S_2 \, S_3
   \, S \, M}^i_{M_2} \right\} \\ 
   & \quad + \left\{\ket{\mu_3 \, S_1 \, S_2 \, S_3 \, S \,
   M}^i_{M_3} \right\} + \left\{\ket{\gamma \, S_1 \, S_2 \, S_3
   \, S \, M}^i_\Gamma \right\} 
\end{split} \end{equation}
and subsequent orthonormalization of the incorporated states.

\begin{figure}[ht!]
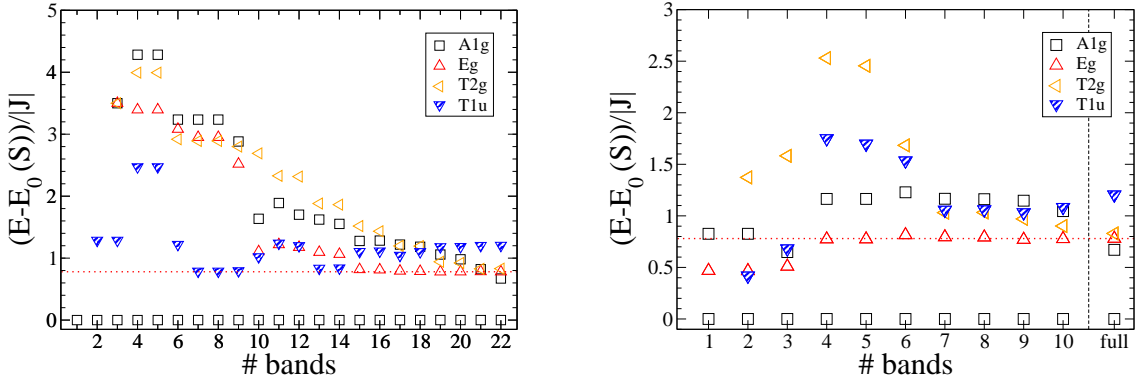
 
   \centering
   \includegraphics*[width=68mm]{ito-frus-fig-6a.eps} \hfill
   \includegraphics*[width=70mm]{ito-frus-fig-6b.eps}
   \caption{Energy difference to the ground state of selected
   low-lying energy levels of the cuboctahedron $s=1$ within
   $S=0$ subspace in dependence on the number of incorporated
   rotational bands. The reduced set of basis states is taken
   from rotational-band states of the $\Gamma(q=0)$-family only
   (left) and from linear combinations of states of all four
   colorings (right). The classification is according to $O_h$
   point-group symmetry. The dotted red line refers to the exact
   difference to the $E_g$-state.}
   \label{F-6}
\end{figure}

In Fig. \ref{F-6} the energy difference between selected
low-lying energy levels and the ground state of the system is
displayed for a cuboctahedron $s=1$ within $S=0$ subspace. The
difference is shown in dependence on the number of incorporated
rotational bands. The last column refers in both subfigures to
the exact values, taken from a complete diagonalization. The
dotted  red line indicates the energy difference to the
$E_g$-state. 

In the first case (left subfigure) only rotational-band states
of the $\Gamma(q=0)$-family are taken into account for the
approximate diagonalization. In the second case (right
subfigure) linear combinations of states of all four colorings
are used. The first noticeable difference is that when taking
basis states of all colorings into account the low-lying levels
start and remain in close proximity to their final (true)
value. This is especially obvious when comparing the convergence
of the lowest $A_{1g}$-state in both subfigures. The second
difference is that when using linear combinations of states of
all four colorings the convergence is smoother and more
rapidly. This can be traced back to the fact that the different
colorings contribute low-lying states of different irreducible
representations which in the other families would only be
available as high-lying states. Therefore, it is advantageous to
use fewer bands but basis states of all classical colorings for
the approximate diagonalization.

A short explanation might be appropriate in order to understand
why the energy differences sometimes increase when taking more
states into account. This is due to the fact that the ground
state in such cases converges more rapidly than the excited
states. Although all approximate eigenvalues improve on an
absolute scale the differences get worse for a moment.

\subsection{Results}

\begin{figure}[ht!]
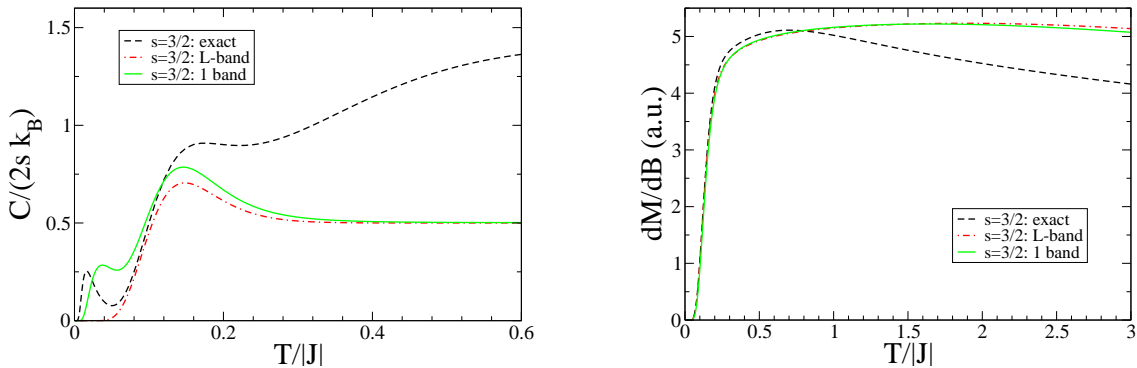
 
   \centering
   \includegraphics*[width=70mm]{ito-frus-fig-7a.eps} \hfill
   \includegraphics*[width=66mm]{ito-frus-fig-7b.eps}
   \caption{Comparison of approximate and exact specific heat
   $C(T,B)$ (left) and zero-field susceptibility $dM/dB$ (right)
   for a cuboctahedron $s=3/2$. The L-band refers to magnetic
   properties evaluated with the lowest band of the rotational-band
   Hamiltonian. For the approximate diagonalization only
   the states of the lowest rotational band but of each
   coloring are taken into account.}
   \label{F-7}
\end{figure}

In this section we like to present how good thermodynamic
observables can be approximated. Figure \ref{F-7} shows the
specific heat $C(T,B)$ (left) and the magnetic susceptibility
$dM/dB$ (right) for zero field $B=0$ of a cuboctahedron
$s=3/2$. The energy spectrum of this system can be completely
calculated using irreducible tensor operator technique in
combination with a $D_2$ point-group symmetry \cite{ScS:P09}. In
Fig. \ref{F-7} the exactly calculated specific heat and the
susceptibility are compared with results from an approximate
diagonalization using only states of the lowest rotational band
in each coloring in order to set up the basis. We also show how
these observables look like if evaluated with only the lowest
rotational band (L-band) of $\op{H}_\text{RB}$ (red colored
energy levels in Fig.~\ref{F-2}).

As one can see the low-temperature specific heat is very
sensitive to the structure of low-lying levels. In the
rotational-band model the lowest band (L-band) is gaped and
describes a rotor. Therefore, the specific heat is suppressed at
small temperatures and at higher temperatures approaches
$3/2$~k$_B$. The approximate diagonalization, although only
taking the lowest (L-band) states of all four coloring into
account, achieves already an improvement for
low-temperatures. The little Schottky-peak is the result of
low-lying singlets and rearranging triplets. At higher
temperatures this approximation displays the same behavior as
the pure L-band which is expected since only L-band like states
are incorporated. 

In contrast to the specific heat which directly reflects the
density of states within a certain energy interval the magnetic
susceptibility reflects the density of magnetic states. To that
end it is not surprising that the magnetic susceptibilities
calculated from the approximate spectra do not differ
considerably. Since the exact spectrum also possesses a
rotational-band like behaviour at the lower edge the
contributions from the approximate diagonalization and from the
L-band reproduce the steep rise of the susceptibility for low
temperatures very well. Obviously, the exact thermodynamical
properties cannot be reproduced properly in the case of
increasing temperature because the approximate spectra only
contain a fraction of the energy levels of the full Hilbert
space $\mathcal{H}$.

\subsection{Approximate selection rule} \label{sec-selection}

As mentioned before (Sec. \ref{sec-convergence}) the stepwise
convergence behaviour leads to an approximate selection
rule. Using this selection rule the computational effort when
setting up the Hamilton matrices can be further reduced.  In
Fig. \ref{F-5} it was shown that when diagonalizing in the
$\Gamma(q=0)$-family, regions can be marked in which single
energy levels are affected by taking into account additional
rotational-band states.  The same can be done for the
$M$-family.  As it was already mentioned, the marked regions
coincide with the incorporation of rotational-band states
belonging to the same super-band.

Looking at Fig. \ref{F-5} the approximate energy eigenvalues of
the ground state are lowered whenever states $\ket{ \gamma \,
S_1 \, S_2 \, S_3 \, S\, M}$ are additionally included which
belong to a super-band with an even number of magnons. Since the
convergence behaviour is state-sensitive the energy of the
lowest states belonging to the 3-dimensional irreducible
representation $T_{1u}$, for example, is considerably lowered
whenever super-bands with an odd number of magnons are
included. This relation between the incorporation of super-bands
and the immediate affection on the energies of certain
approximate eigenstates can also be observed if the classical
ground state belongs to the $M$-family.

As a result of the aforementioned observations a simple rule can
be conjectured for low-lying energy levels $\ket{S_{1,b} \,
S_{2,b} \, S_{3,b} \, S \, M}$. The matrix elements
$\bra{S_{1,a} \, S_{2,a} \, S_{3,a} \, S \, M}\op{H}\ket{S_{1,b}
\, S_{2,b} \, S_{3,b} \, S \, M}$ connecting these states with
other states $\ket{S_{1,a} \, S_{2,a} \, S_{3,a} \, S \, M}$ are
one magnitude (or more) bigger than other matrix elements if
\begin{equation} \label{eq:selrule}
   |n_a - n_b| \mod 2 = 0
   \ ,
\end{equation}
where $n_a=\sum_{i=1}^3 S_{i,\text{max}} - \sum_{i=1}^3 S_{i,a}$
and $n_b$ (similarly evaluated), denote the number of magnons of
the super-bands the states are belonging to. The Hamilton
matrices approximately split up according to
Eq. \fmref{eq:selrule}. Thus, simultaneously computation time is
saved and the dimensionality of the problem is reduced.

\section{Summary} \label{sec-4}

In this paper an approximate diagonalization scheme was proposed
and used in order to determine the energy spectra of
geometrically frustrated spin systems. As an example the
cuboctahedron was discussed. It was shown that the convergence
is clearly dependent on the coloring of the underlying classical
ground state, i.e on the coupling scheme. Therefore, the
approximation can be improved by using an adapted set of basis
states that originates from rotational-band states of all possible
sublattice colorings. Furthermore, it was shown that the
approximate diagonalization is useful in order to study the
low-temperature thermodynamics of geometrically frustrated
systems, even in its simplest form when only states from one
rotational band within each coloring are taken into account.

Although the necessary calculations are rather involved,
especially for huge frustrated systems, the approximate
diagonalization can be a valuable method. Numerically the
evaluation of recoupling coefficients constitutes the strongest
challenge, i.e. calculations are rather limited by runtime than
by available memory. However, recent developments show that
highly parallelized program code or public resource computing
can help to overcome this barrier.

\section*{Acknowledgements}{Computing time at the Leibniz
  Computing Centre in Garching is greatfully acknowledged as
  well as helpful advices of Mohammed Allalen. We also thank
  Boris Tsukerblat for fruitful discussions about the
  irreducible tensor operator technique. This work was supported
  within a Ph.D. program of the State of Lower Saxony in
  Osnabr\"uck.} 


\end{document}